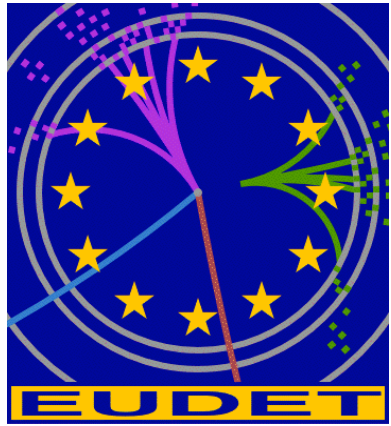

# Beam test of the QMB6 calibration board and HBU0 prototype

J. Cvach[1], J. Kvasnička[1,2], I. Polák[1], J. Zálešák[1]

May 23, 2011

## Abstract

We report about the performance of the HBU0 board and the optical calibration system QMB6 in the DESY test beam. A MIP signal was measured on the HBU0 board, providing a reference to the LED amplitude scan, showing an amount of light delivered to each of 12 SiPMs from 0 to 250 MIP at maximum. An ASIC analogue memory cell performance was analyzed in relation with the SiPM signal strength. We observed a non-uniformity and a pedestal shift in the ASIC readout values in the high gain mode for the higher signals strength.

---

[1] Institute of Physics of the ASCR, v.v.i., Prague, Czech Republic
[2] Czech Technical University, Prague, Czech Republic

EUDET-Memo-2010-21

# 1 Introduction

The test was performed at DESY Hamburg at test beam area 21. It was organized to show a common performance of two boards: the HBU0 and QMB6.

## 1.1 HBU0

The HBU0 (HCAL Base Unit)[1] is the smallest compact unit of the engineering prototype of the CALICE high-granularity hadron calorimeter, containing 144 pieces of scintillator tiles which have dimensions of 30×30×3 mm$^3$ each. The board has a total size of 36cm×36cm (12×12 scintillator tiles). Each tile is read out with one SiPM and one ASIC on the board reads out 36 SiPMs.

## 1.2 QMB6

The QMB6 [3] is a prototype board containing 6 channels of the Quasi-resonant LED driver [2]. The LED driver generates an optical pulse using the UV-LED (Ultra-violet LED). The pulse width is fixed by the PCB design. The amplitude is steerable from zero to the full range via a Labview control. Each LED is connected to one notched optical fiber which distributes the light equally (within 20% of deviation) among 12 positions of tiles in a row.

### 1.2.1 Controlling the QMB6

The light output from the QMB6 is controlled by 2 parameters: V1 and V2. These parameters correspond to control voltages of the Quasi-resonant LED driver. In principle, the V2 should be higher than V1 (safe margin is when V1>V2-1000). The light output increases with increasing V1 and decreasing V2. For the full range, V2 can be fixed at some value (3095) and the light output can be steered just by V1 setting (0~4095).The higher V1, the higher light output.

The QMB6 is controlled via the Labview program (shown at Figure 1), which controls the CANBUS interface, through which are commands sent to the QMB6 board. The QMB6 is capable of readout of the board temperature and all important voltages.

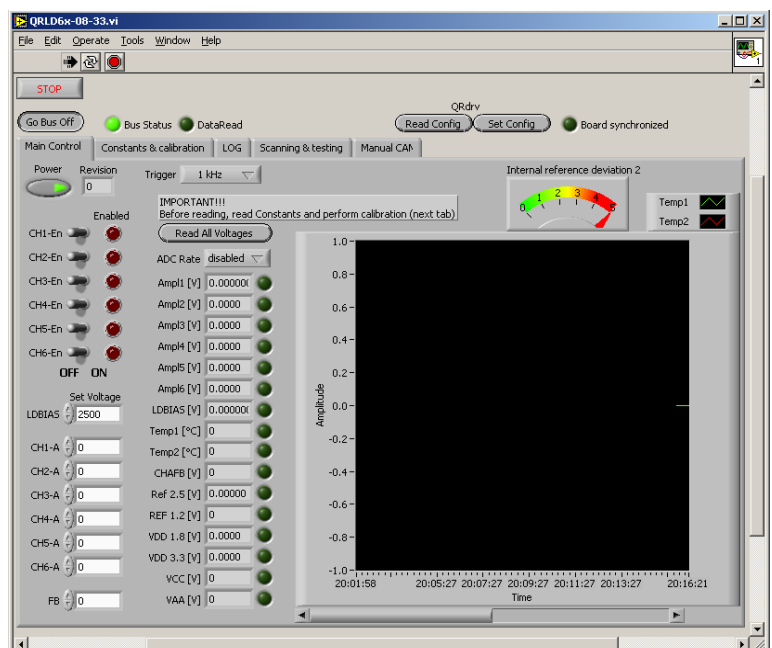

Figure 1: Labview control of the QMB6

## 1.3 Beam setup

The measurement was performed at the test beam area 21 during the period from July 19 to July 30, 2010. Data was taken in 2 different configurations:
- MIP measurement: with beam on and external beam coincidence trigger
- LED flashing response: beam off, triggered internally by the CALIB board sitting on the top of the HBU0

The test setup of the measurement is shown in Figure 2. The beam coming from the beam magnet comes at first to the coincidence of PMTs. Signals from the PMTs are processed in the VME crate, creating a beam gate and event trigger signals to the HBU0.

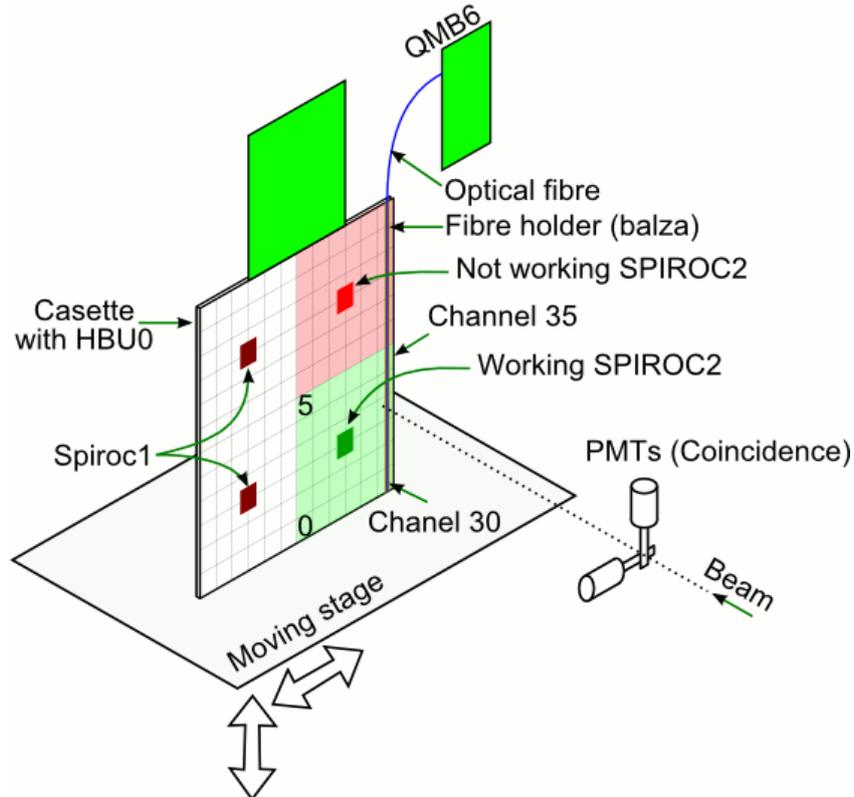

**Figure 2: Test beam setup**

The HBU0 and QMB6 were fixed at the moving stage. The beam irradiated 1 scintillator tile at the time. The tile was selected by the positioning of the moving stage to the beam.
The QMB6 was placed above the HBU0 electronics. The optical signal from the QMB6 was delivered to the HBU0 via 1 fiber (Figure 2), which was connected to one of the LEDs at the QMB6. The fiber was kept in the straight position by a 5mm thin balsa wood rod (Figure 3).
The balsa wood has holes at positions where are the notches of the fiber. These holes were used for the alignment to match the notches and the scintillator alignment pins.

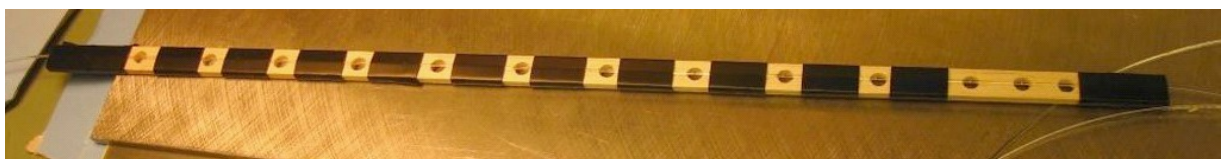

**Figure 3: Optical fiber support**

The fibre was fixed at the bottom row of scintillator tiles of the HBU0 (Figure 2). Light from the notches of the fiber illuminated the HBU0's tiles through the scintillator tile alignment

EUDET-Memo-2010-21

pins (Figure 4). Illuminated channels of the HBU0 were channels no. 30~35 of each of the SPIROC2 chip. This setup gives 12 illuminated tiles by a single fiber but only 6 tiles with working readout.

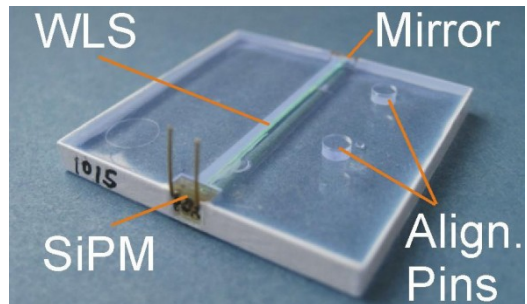
**Figure 4: Scintillator tile and alignment pins**

## 2 MIP signal

Measurement of the MIP signal is a method how to calibrate the tile and SiPM response to a defined energy. The beam energy remains the same during the whole measurement, therefore 1) channels throughout the HBU0 can be equalized to the same level; 2) gain of the ASIC can be evaluated within different ASIC configurations (feedback capacitance, gain mode).

### 2.1 MIP measurement

We planned to scan tiles 30~35 of the Spiroc2 number 0 (ASIC0) and Spiroc2 number 1 (ASIC1). Due to fault of the configuration of the ASIC no. 1 on the HBU, we were able to make measurements only for 6 tiles (no. 30~35).
Statistics for this measurement was 10000 events in each ASIC memory cell (= full memory pipeline). SPIROC2 was configured for the highest sensitivity (100fF feedback capacitance, high gain mode).
Two of 16 memory cells are shown at the Figure 5, as indicated at the label of each histogram. The event 1 (right column) is the first stored event, since the event 0 is not read out due to an ASIC bug. The event 15 (left column) is the last recorded event.
Channel 34 showed no signal; therefore we had only 5 working channels. We saw a beautiful single photon-electron (PE) spectrum at channel 31, as shown at Figure 5. Other channels do not show so clear single PE structure (channels 30 and 32) in the MIP signal, or do not show the structure at all (channels 33 and 35).
The "pedestal" values shown in Figure 5 (green histogram) were measured when a beam was positioned onto a neighbor cell. This position shows a nice pedestal with the $1^{st}$ pixel.
After a measurement in high gain we changed the ASIC configuration and we measured in low gain (with the feedback capacitance of 100 fF) with the same beam position.

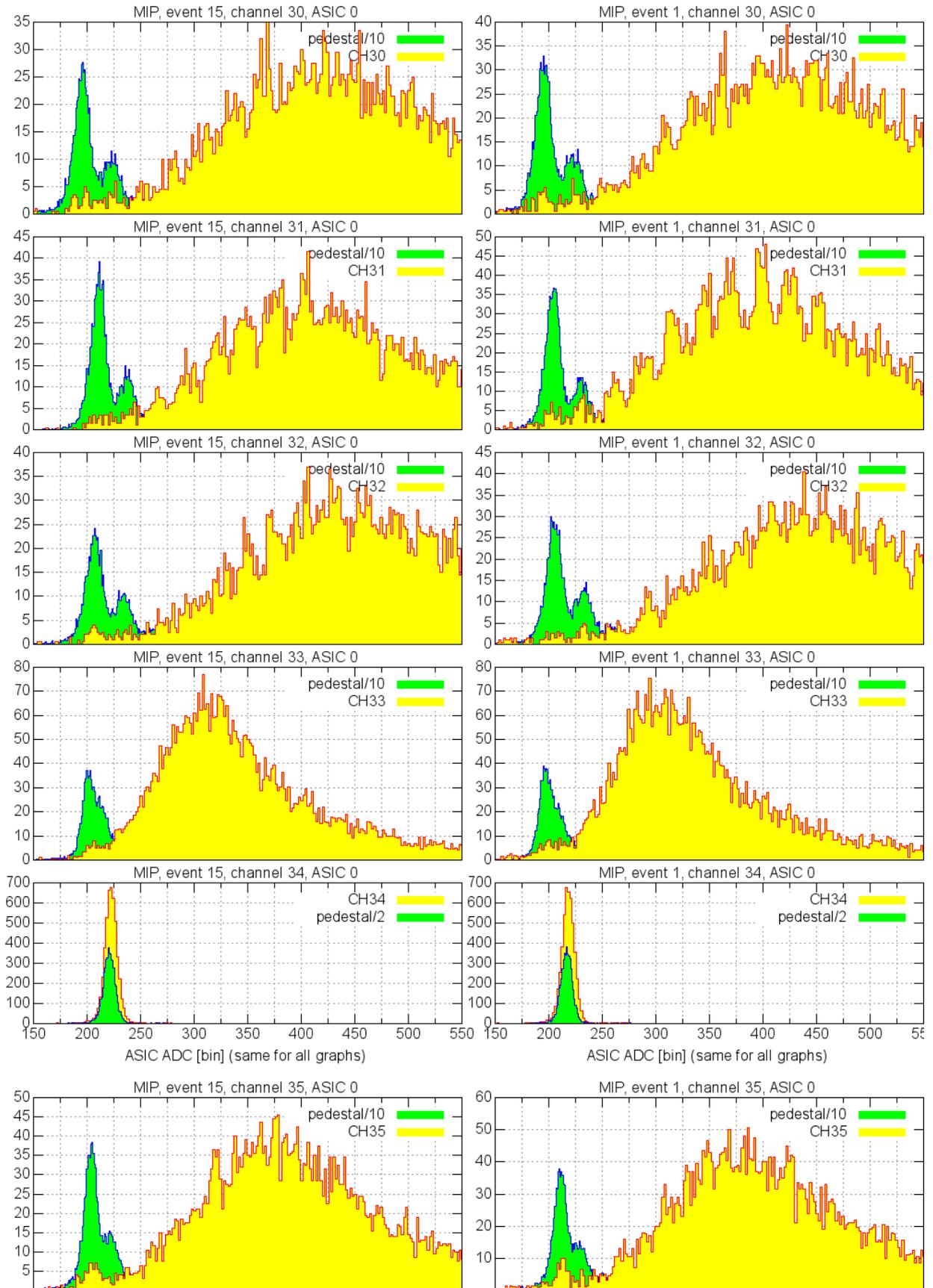

**Figure 5:** Beam MIP single photon-electron spectrum. Left column: last event in the ASIC memory cell array. Right column: first event in the ASIC memory cell array. The X axis scale is the same for all figures

## 3 Gain calibration

The gain between HG and LG is an important factor for comparing result between high gain mode (which is used for calibration and also for MIP measurement) and HG mode (mode for taking the data).

The ratio between the HG and LG was calculated for each memory cell and evaluated individually accordingly to the formula:

Ratio = (HG_MIP_signal – HG_pedestal)/(LG_MIP_signal – LG_pedestal).

The results are shown in Figure 6 and Figure 7. These figures show a non-uniformity of the ASIC gain within the memory cells and among the channels. In an ideal case: 1) each memory cell within one channel would have the same gain ratio; 2) all ASIC channels would have same (fixed) gain ratio.

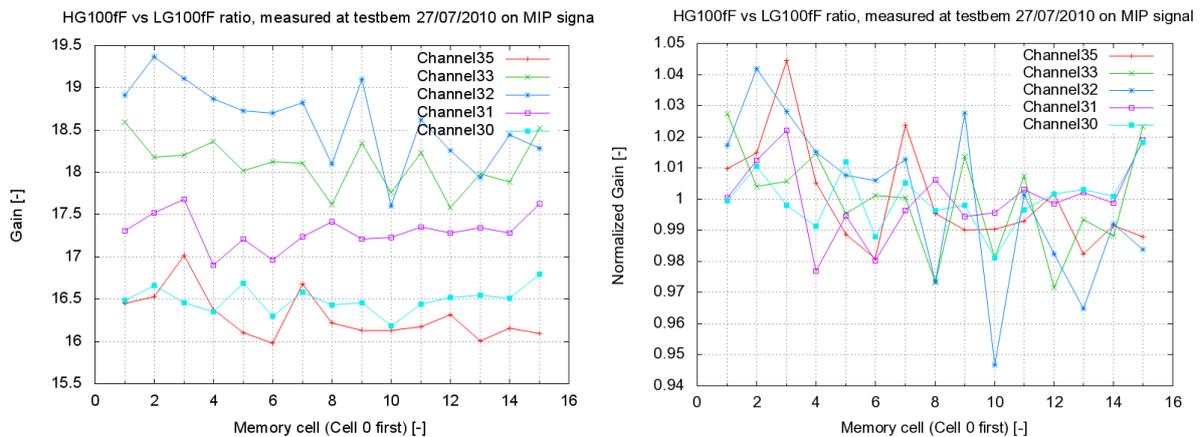

**Figure 6: absolute (left) and normalized (right) ratio between HG and LG in the memory cells. Reference for the normalized gain is the mean value throughout all memory cells**

Gain ratios displayed in Figure 7 vary from 16.3 to 18.6, with a deviation of 1 to 2.5 %.

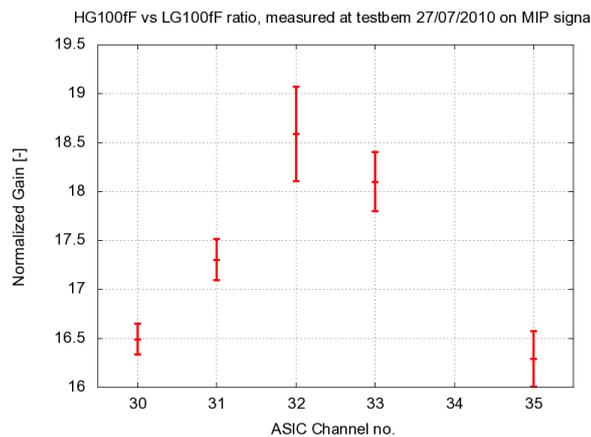

**Figure 7: Mean value and standard deviation of the ASIC channels**

## 4 Amplitude scan

We scanned the HBU response by the QMB6 board to a full range of operation: from 0 (no light) to maximum light output (deep ASIC ADC saturation). The QMB6 is steered by a V1 DAC setting, which sets up a light intensity coming from the LED. The second QMB6 control voltage (V2) was fixed at the value of 3095 bins. We observed several effects during the measurement.

## 4.1 Pedestal shift due to saturation

The pedestal shift is 30 ADC bins in the worst case (equivalent to approx. 1 pixel hit) for the LED signal at 80% of the ASIC ADC saturation level. When the LED signal is at 30% of saturation level, the pedestal is shifted by 10 ADC bins in the worst case. This gives a pedestal shift of 1.0% to 1.2%, assuming all signals have the same level. Full overview of the amplitude scan is shown in Figure 8 and Figure 9 in detail, respectively. This effect makes an influence at almost all channels except the dead channel (#3) and channels, which have higher crosstalk (#25 and #28).

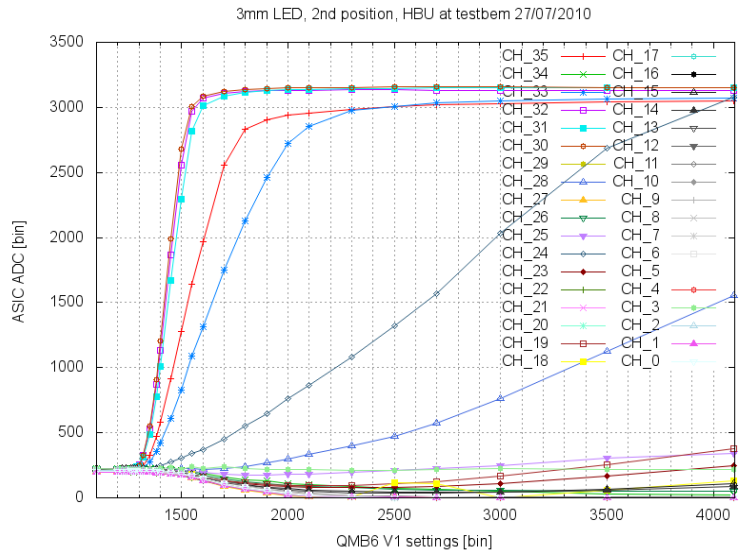

**Figure 8: Amplitude scan in the High gain**

The measurement was performed by a fiber which illuminated 6 tiles simultaneously. We expect results would be probably different in case of only one tile is illuminated.
As a result of this pedestal shift, low signals cannot be measured together with significantly stronger signals. The pedestal shift degrades such measurements.

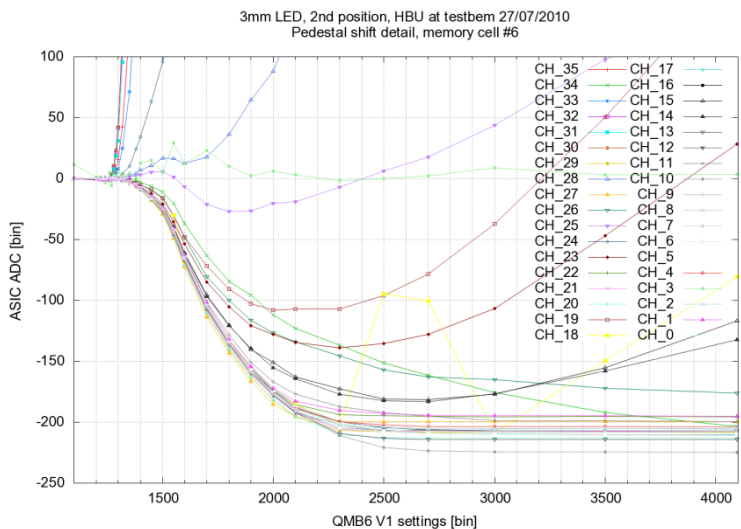

**Figure 9: Detail of the amplitude scan**

## 4.2 Memory cells uniformity

Another observed issue was a signal shift throughout the memory cells. Without any optical signal being applied, the signal (actually the pedestal) is higher for the first 3 events (50 bins higher than the others, equals to 2 pixels hit), than there is a jump to a base level where the pedestal of each memory cell has a different value (within 20 bins range).

Figure 10 gives insight into the problem: rightmost cells (first events) give higher signal than the other memory cells. Reference is the memory cell 0 (latest data, event #15).

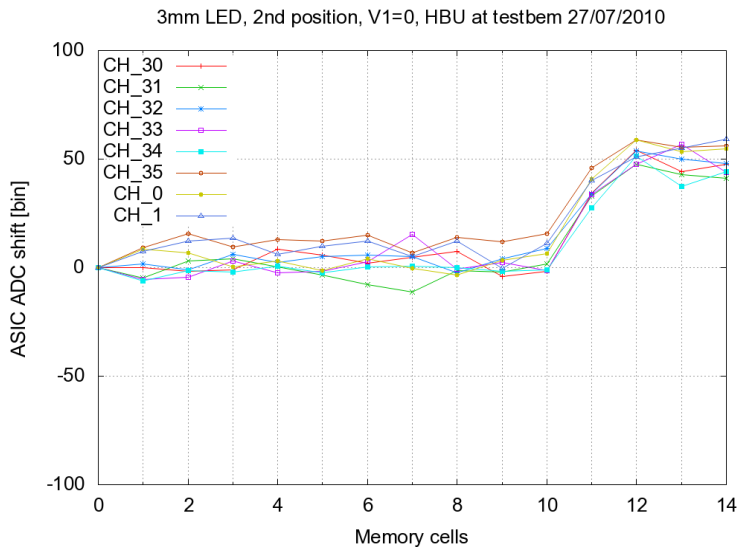

**Figure 10: Different signal at memory cells**

The profile of the memory cells response changes significantly, as the signal amplitude increases. We observed, that the dependence is not possible to describe with a simple function, therefore it is shown at Figure 11 and Figure 12. Signal distribution, when the optical signal reaches 80% of the HG saturation level, is shown in Figure 11. Some illuminated channels have smaller first events (channels 30, 31 and 32 at Figure 11), than the signal increase (first events on the right side, last events are on the left side). Reference value in at Figure 11 and Figure 12 is always the last event from the channel.

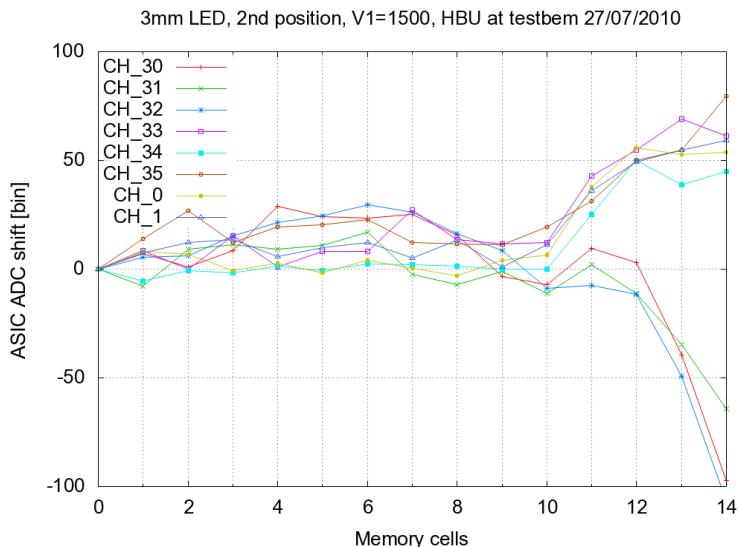

**Figure 11: Signal in memory cells on the verge of saturation**

The shape of the signal in memory cells (Figure 12) changes, as the ASIC saturation gets deeper (V1=2000, Figure 12 left) to a maximum level (V1=4095, Figure 12 right), where the pedestal of non-illuminated channels (channels 0 and 1) reaches the lowest level of the ADC of SPIROC2, therefore the non-illuminated channels (channels 0 and 1) show constant 0, except for the first event (memory cell 14) .

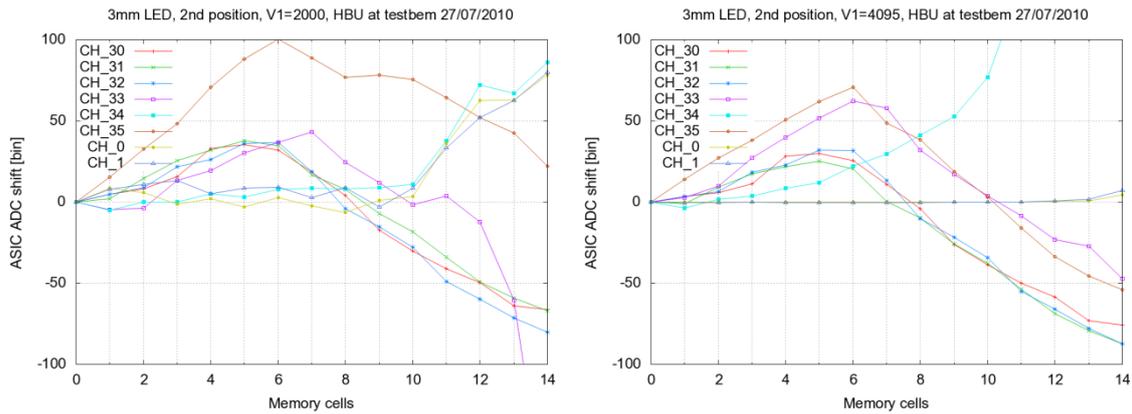

**Figure 12: Distribution of signal among the Memory Cells during a the saturation**

The complex variation of the signal over the memory cells and optical signal is shown at Figure 13 and Figure 14. These figures show an absolute ADC readout value on the Z-axis and a difference from the last event on the color scale.

Figure 13 show one of the illuminated channels, channel 35. When the light was turned off (V1=0 cut), a nice pedestal with first 4 events shift is shown. This pedestal shift is similar to Figure 10. The ADC gets saturated near V1=1600 bins, the key message of this graph is the change of the Memory cells profile as the light signal increases.

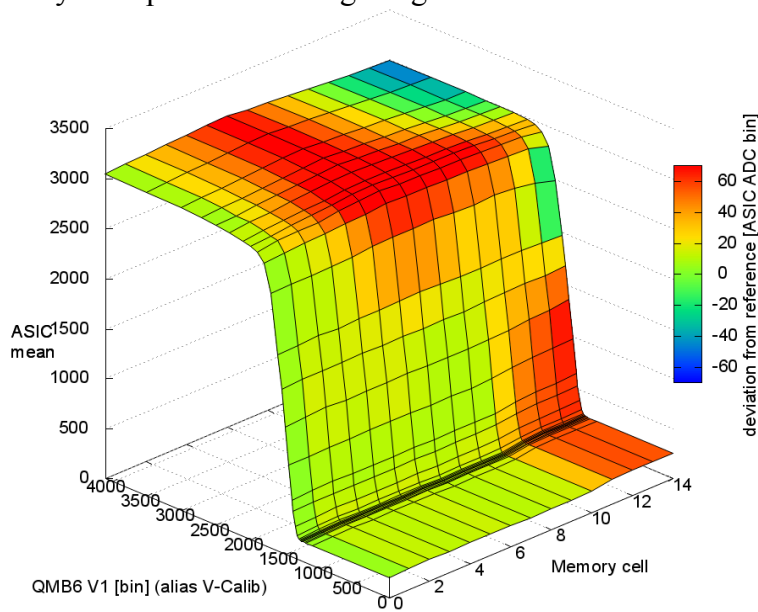

**Figure 13: Amplitude scan in the memory cells of channel 35**

Figure 14 shows a shift of the non-illuminated channel #1 during the illumination of channels 30~35 (shown at Figure 13). Selections from Figure 13 are shown at Figure 9 (memory cell is fixed) and Figure 11 (V1 setting is fixed).

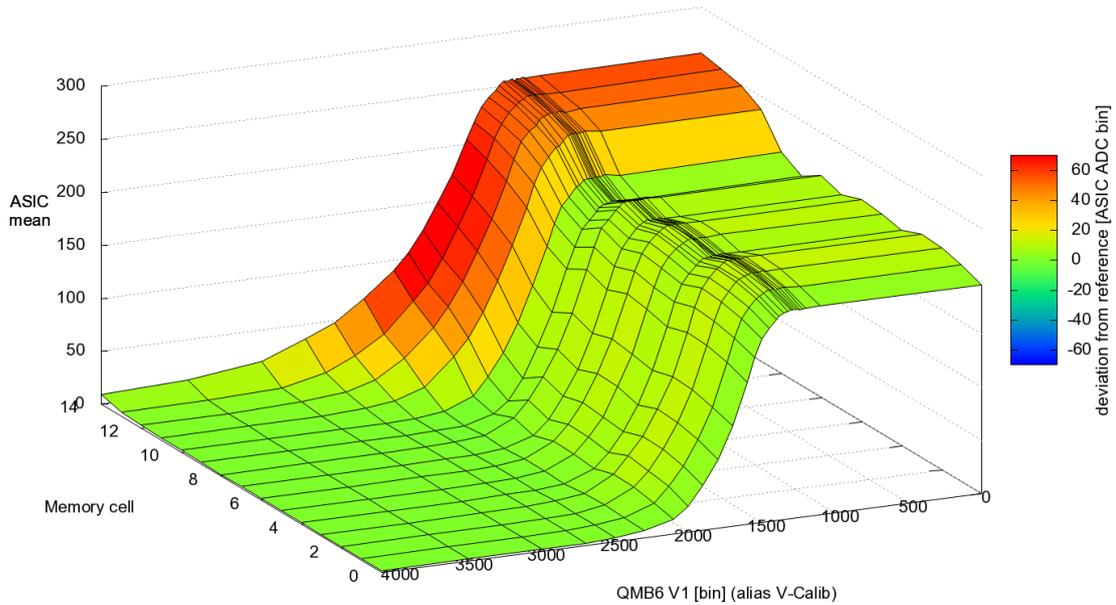

**Figure 14: Signal shift at the non-illuminated cell #1 during illumination of channels 30~35**

The shifts within the memory cells seem to have a common structure, which is preserved during all V1 setting and which remains invariant. This leads to an idea to subtract the memory cell structure, which is shown at Figure 15. Only the most interesting range (no light up to ASIC saturation at V1=1600) is shown at Figure 15 and the color range was rescaled to <-5,5> bins. After the subtraction, only small residuals ~ ±1 bin remain, except for memory cells 12 and 11 (3$^{rd}$ and 4$^{th}$ event), which have higher drop (emphasizes by a blue color at Figure 15).

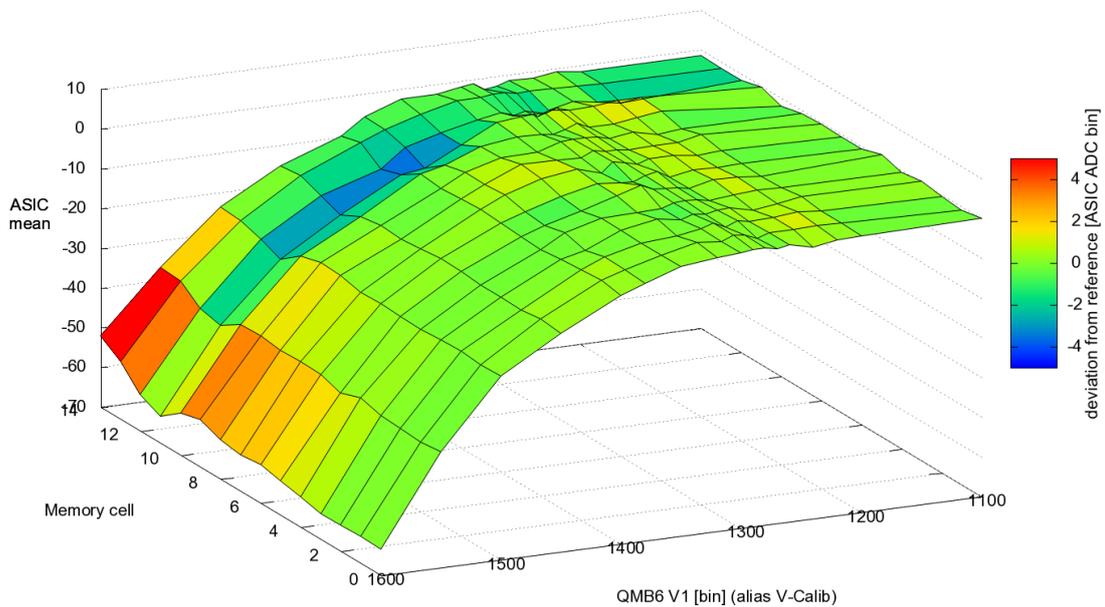

**Figure 15: Detail of the pedestal shift for non-illuminated channel (#1), when inter-memory-cell shift is compensated and memory cell 0 (last event) is subtracted as a reference value.**

EUDET-Memo-2010-21### 4.3 Amplitude scan in the Low gain mode and QMB6 efficiency

We evaluated the maximal light delivered in the MIP scale. This measurement was performed before the testbeam, at April 2010 at DESY Hamburg in the electronics laboratory. The estimation was preceded by

1) optical power measurements of the QMB6 (to get rid of the non-linear V1 scale and to replace it by a *linear* optical power output scale);
2) measurement of the distance among the single photon-electron peaks in HG mode;
3) evaluation of the HG/LG ratio of the ASIC;
4) MIP measurement;

From these measurements we got a saturation curve (Figure 16). The initial slope of the saturation curve ($f(x) = x \cdot 5.5 \cdot 10^{12}$, where $x$ is the optical pulse energy and $f$ is extrapolated number of fired pixels) is not affected by saturation and was used to evaluate the maximal light delivered to the tile with SiPM through the alignment pin of the scintillator tile. The maximum light power corresponds to 2500 pixels hit for the maximal optical power ($4.6 \cdot 10^{-10}$ J), which represents 255 MIPs with a high uncertainty due to an unstable optical coupling between the LED and optical fiber and due to the spread of the light output from the notches (15% spread among the notches).

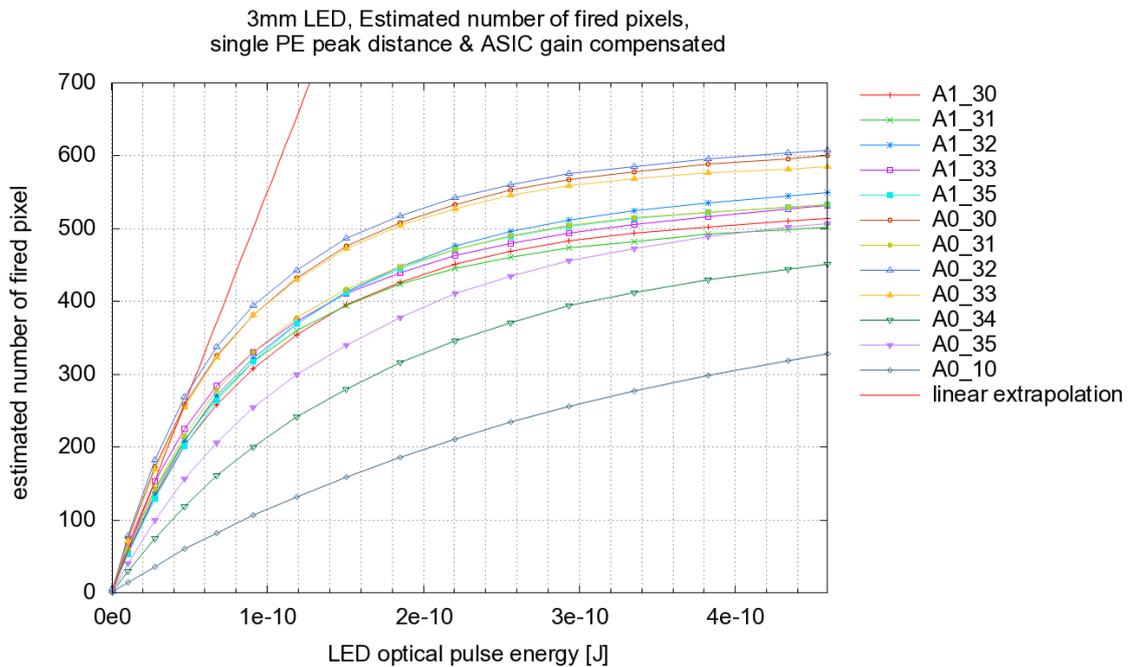

**Figure 16: Linear approximation of the initial slope of the saturation curve of the SiPM**

## 5 Summary

The combined beam test of QMB6 and HBU0 provided the first possibility to operate together in real experimental conditions both systems. The beam signal provided a reference scale for the QMB6 LED amplitude in MIPs. The dynamic range of the QMB6 LED light amplitude ranges from 0 to 250 MIPs. The analysis of test beam data allowed us to investigate the internal properties of the ASIC. The results are important for further improvements of the next generation of the ISIC prototypes.

- 11 -

## Acknowledgement

This work is supported by the Commission of the European Communities under the 6$^{th}$ Framework Programme "Structuring the European Research Area", contract number RII3-026126 and by the Ministry of Education, Youth and Sport of the Czech Republic under the projects LC527 and LA09042.